\documentclass[conference,letterpaper]{IEEEtran}
\usepackage[letterpaper, left=0.76in, right=0.76in, bottom=0.77in, top=0.95in]{geometry}
\IEEEoverridecommandlockouts
\usepackage{cite}
\usepackage[utf8]{inputenc} 
\usepackage{amsmath,amssymb,amsfonts}
\usepackage{algorithmic}
\usepackage{graphicx}
\usepackage{url}
\usepackage{textcomp}
\usepackage{xcolor}
\usepackage{etoolbox}
\usepackage{tabularx,booktabs}
\usepackage{float}
\def\BibTeX{{\rm B\kern-.05em{\sc i\kern-.025em b}\kern-.08em
    T\kern-.1667em\lower.7ex\hbox{E}\kern-.125emX}}

\newcommand{\tech}[0]{AI-CDA4ALL }

\begin{document}

\title{AI-CDA4All: Democratizing Cooperative Autonomous Driving for All Drivers via Affordable Dash-cam Hardware and Open-source AI Software}




\author{\IEEEauthorblockN{ Shengming Yuan}
\IEEEauthorblockA{\textit{Department of Civil and Environmental Engineering} \\
\textit{University of South Florida}\\
Tampa, United States \\
shengming@usf.edu}
\and 
\IEEEauthorblockN{ Hao Zhou*\thanks{\,Hao Zhou is the corresponding author (email: \texttt{haozhou1@usf.edu}).}}
\IEEEauthorblockA{\textit{Department of Civil and Environmental Engineering} \\
\textit{University of South Florida}\\
Tampa, United States \\
haozhou1@usf.edu}
}

\maketitle

\begin{abstract}

As transportation technology advances, the demand for connected vehicle infrastructure has greatly increased to improve their efficiency and safety. One area of advancement, Cooperative Driving Automation (CDA) still relies on expensive autonomy sensors or connectivity units and are not interoperable across existing market car makes/models, limiting its scalability on public roads. To fill these gaps, this paper presents a novel approach to democratizing CDA technology, it leverages low-cost, commercially available edge devices such as vehicle dash-cams and open-source software to make the technology accessible and scalable to be used in transportation infrastructure and broader public domains. This study also investigates the feasibility of utilizing cost-effective communication protocols based on LTE and WiFi. These technologies enable lightweight Vehicle-to-Everything (V2X) communications, facilitating real-time data exchange between vehicles and infrastructure. Our research and development efforts are aligned with industrial standards to ensure compatibility and future integration into existing transportation ecosystems. By prioritizing infrastructure-oriented applications, such as improved traffic flow management, this approach seeks to deliver tangible societal benefits without directly competing with vehicle OEMs. As recent advancement of Generative AI (GenAI), there is no standardized integration of GenAI technologies into open-source CDAs, as the current trends of muiltimodal large language models gain popularity, we demonstrated a feasible locally deployed edge LLM models can enhance driving experience while preserving privacy and security compared to cloud-connected solutions. The proposed system underscores the potential of low-cost, scalable solutions in advancing CDA functionality, paving the way for smarter, safer, and more inclusive transportation networks.
\end{abstract}


\begin{figure*}
	\centering
	\includegraphics[width=1\linewidth]{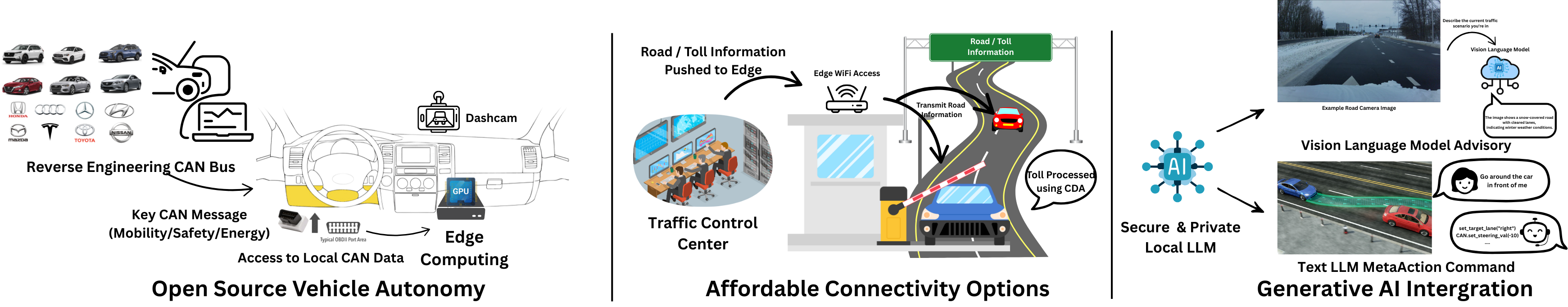}
	\caption{\tech Ecosystem}
	\label{fig:tech_ecosystem}
\end{figure*}

\section{Introduction}

Autonomous driving technologies have seen rapid advancements in recent years, yet they remain largely inaccessible to the general public. Most commercial systems are built on expensive sensor suites, proprietary algorithms, and closed vehicle platforms, making them cost-prohibitive and non-interoperable across different car makes and models. This is particularly limiting given that, since 2016, the majority of production vehicles have adopted electronic controlled, digitally manage core driving functions such as steering, braking, and acceleration. Through community-driven reverse engineering, these control interfaces have been unlocked across hundreds of car models\cite{opendbc-git}, revealing that low-cost, software-based autonomy is technically feasible using readily available consumer hardware—such as a smartphone or dash-mounted camera—as the primary computation and sensing platform.

In parallel, open-source self-driving frameworks such as OpenPilot\cite{openpilot-git}, Autoware\cite{kato2018autoware}, and others have significantly lowered the barrier to autonomous development. These platforms enable individuals and researchers to design and test perception, planning, and control modules on real vehicles, offering high customizability and extensibility. Furthermore, they present a powerful opportunity for infrastructure stakeholders—such as cities or transportation agencies—to build advisory or cooperative driving applications directly into the vehicle software stack. While community developers have embraced these systems for personal projects, the research community and public infrastructure stakeholders have yet to fully engage. Many academic innovations remain confined to in-house testbeds and fail to transition into scalable, real-world deployments. Infrastructure owners, in turn, face barriers in interfacing with proprietary OEM systems. This disconnect creates a bottleneck between research outcomes and real-world impact. We argue that a more open and participatory ecosystem is needed—one that empowers customers, researchers, and infrastructure stakeholders to collaboratively build and improve connected autonomy systems.

A critical component of cooperative driving is robust vehicle connectivity. Yet most high-performance communication solutions—such as PC5\cite{s21030843} used in advanced V2X solutions, require dedicated and expensive onboard units, severely limiting scalability. Cellular and WiFi-based alternatives have historically been dismissed due to performance limitations, but the landscape has changed: modern protocols like LTE-Advanced, 5G, and WiFi 6 offer significantly improved reliability and throughput \cite{fi14100293}. These advances make them promising candidates for non-safety-critical V2I and V2X communications, such as traffic advisory messages, infrastructure updates, and digital twin interactions. In this work, we revisit the viability of these modern wireless protocols for connected driving, evaluating their performance and integrating them into the open-source self-driving stack. To support broader use, we provide standardized APIs for connectivity-based functions and enable infrastructure applications to interface with vehicles directly, reducing deployment friction.

Simultaneously, the rapid emergence of generative AI (GenAI) models, particularly multimodal large language models, offers exciting potential to enhance driving autonomy through reasoning, personalization, and decision support. Despite early explorations in the literature \cite{tang2025autoagentfullyautomatedzerocodeframework}, GenAI integration into real-time driving and connectivity systems remains limited. A key obstacle is the absence of standardized interfaces that link GenAI capabilities with widely used open-source self-driving stacks. To address this, we build a lightweight API layer that connects GenAI models with both driving logic and communication modules. In addition, we introduce meta-action functions based on the internal design of the baseline self-driving system, allowing GenAI models to translate abstract reasoning into executable commands more accurately and efficiently.

\textbf{In this paper}, we aim to address these critical research gaps by building an open-source, affordable, and standardized CDA system with integrated AI capabilities. Our goal is to democratize AI-enabled CDA across three key communities: everyday drivers, academic researchers, and infrastructure stakeholders. By enabling scalable real-world deployments, fostering interoperability, and supporting intelligent interactions between vehicles and infrastructure, we seek to accelerate the development of a more open, inclusive, and collaborative ecosystem for connected and autonomous transportation.


Motivated by the aforementioned research gap and guided by a \textbf{transportation engineering perspective}, this paper contributes to the following main efforts:

\begin{enumerate}
    \item \textbf{Scalable and Affordable Open-source Autonomy}:
    We build upon a widely adopted Level-2 open-source self-driving platform to ensure affordability, extensive compatibility with existing car models, and a straightforward plug-and-play experience. We have tested our platform extensively on over 10 different car make and model.

    \item \textbf{Lightweight Connectivity Development and Integration}:
    We proposed, implemented and demonstrated a lightweight, standards-compliant connectivity system that adheres to SAE V2X standards J2735\cite{j2735}, which is able to transmit non-safety-critical ones (e.g., toll information) using low cost protocol over WiFi and LTE connections. Transmitting message such as advisory speed messages to alleviate congestion, enhance road safety, and enable cooperative driving (e.g., queue formation and merging assistance). 

    \item \textbf{API Development for GenAI Integration}:
    We design a suite of custom APIs (i.e. MetaAction) that bridge the gap between high-level LLM outputs and executable commands for scalable open-source autonomy. Through code-generation mechanisms, Large Language Models (LLMs) produce structured, valid code segments that can be parsed and executed, thus providing a robust method for incorporating GenAI-based decision-making into real on-road scenarios. We provided a foundational development ground for any future Generative AI platform to integrate to \tech.
\end{enumerate}

\section{Related Work}

\textbf{Accessibility of Current Autonomous Driving}:
Many existing research papers on Connected and Automated Vehicles (CAVs) have demonstrated their solutions on one or two specific car models, typically using proprietary hardware or manufacturer-specific control systems \cite{Mehr_2023, info16040317}. These implementations, while valuable as proofs of concept, lack scalability and interoperability, making them challenging to deploy across a wider range of vehicles. This pose a significant gap between research community and industry which both, does not use open source system and cannot transfer technology to each other. In contrast, our approach leverages OpenDBC\cite{opendbc-git} (Open Database of Car), an open-source project that decodes and standardizes CAN bus protocols across numerous car brands and models, allowing us to port our solution to a much broader array of vehicles.
    
\textbf{Current Connectivity Hardware Adoption}:
Despite previous studies that have found many beneficial factors associated with connected driving automation using V2X or V2V technology such as Eco-driving and routing \cite{Othman2022}. Many studies omit the actual implementation of real-world hardware \cite{9304556, KATRAKAZAS2015416}  or are using expensive connectivity hardware meant for research testing, further slowing down the widespread adoption. Most current connectivity studies \cite{s21030843, s18051527} lean heavily on PC5 for safety applications, such as collision warnings and pedestrian detection, due to its direct communication capabilities. Connectivity also facilitates real-time data exchange for maintaining vehicle spacing and coordination, which can facilitate vehicle platooning control. 

\textbf{Current GenAI Autonomy Integration}:
The surge in Large Language Model (LLM) popularity has spurred efforts to integrate Generative AI into autonomous driving, leveraging models like ChatGPT for tasks such as motion planning and visual-language processing \cite{mao2023gptdriverlearningdrivegpt, wen2023roadgpt4visionearlyexplorations}. Studies like Yang et al. (2024) highlight LLMs’ potential in enhancing open-world understanding and decision-making, yet scaling these implementations faces challenges, including high latency critical for real-time decisions, the opaque nature of models complicating validation, and risks of hallucinations leading to unsafe outputs \cite{yang2024llm4drivesurveylargelanguage}. In code generation, frameworks like LangProp \cite{ishida2024langpropcodeoptimizationframework} optimize LLM-generated driving policies, but initial outputs often fail on edge cases, requiring iterative refinement . This reveals a critical gap: the need for structured API functions or meta-actions to guide LLMs in producing accurate, executable, and safe code, as suggested by works on improving code reliability \cite{tang2025autoagentfullyautomatedzerocodeframework}. Developing such tools is essential to bridge this gap, enabling robust GenAI applications in autonomous driving.

\textbf{Current AI Dashcam limitations}:
As dashcam use become widespread among vehicle owners in both commercial and private use settings. Table \ref{commericaldashcam} summarizes a comparison between \tech and commercially available AI-Empowered dash-cam on the market. Although many products have begun to add more autonomy and safety features such as collision warnings and remote vehicle tracker, vendors fail to fully utilize their compute power to enable possible CDA task. \tech demonstrated the capabilities to add advance autonomy with existing dashcam product.

\begin{table*}[htbp]
\centering
\begin{tabular}{|c|c|c|c|c|c|c|}
\hline

System & Sensor Included & Connected & Safety Warning  & Road Users Detection & V2X Enabled & Open Source \\

\hline

Nexar \cite{nexar_2021} & Camera, GPS, IMU & $\checkmark$ & $\times$ \/& $\times$  \/& $\times$ \/& $\times$ \\
\hline

Motive \cite{gomotive.com_2022} & Camera, GPS, IMU & $\checkmark$ & $\checkmark$ \/& $\times$  \/& $\times$ \/& $\times$ \\
\hline

Samsara \cite{samsara} & Camera, GPS, IMU & $\checkmark$  & $\checkmark$ \/& $\times$  \/& $\times$ \/& $\times$ \\
\hline

Vision Connect \cite{connectedwise} & Camera, GPS, IMU & $\checkmark$ & $\checkmark$ \/& $\checkmark$ \/& $\times$ \/& $\times$ \\
\hline

\tech & Camera,C-V2X,GPS, IMU & $\checkmark$ & $\checkmark$ & $\checkmark$ \/& $\checkmark$ \/& $\checkmark$ \\
\hline
\end{tabular}
\caption{Summary of Related Commercially available AI-Empowered Dash-cam}\label{commericaldashcam}
\end{table*}

\textbf{Transportation-oriented AI-CDA Developments}: While CDA systems have made significant strides in advancing vehicle autonomy, they often lack specific design considerations tailored to the unique needs of transportation research and applications. Transportation agencies are eager to embrace new connectivity and AI technologies; however, they frequently encounter challenges due to the absence of user-friendly interfaces and the opaque nature of many AI systems \cite{9564825,xu2023opencdaopensourceecosystemcooperative}. These challenges underscore the need for CDA systems that are not only technologically advanced but also designed with the specific requirements of transportation research and applications in mind. \tech can offer intuitive interfaces and robust control mechanisms to ensure that transportation agencies can effectively leverage CDA technologies while maintaining oversight and accountability.

\section{System Architecture Design}

The \tech system is designed with a focus on integration of affordable, scalable, and open-source components to achieve Cooperative Driving Automation System. The system incorporates low-cost hardware, V2X connectivity, and cutting-edge GenAI integration to offer a flexible, extensible platform that supports real-world deployment. The architecture is designed to be modular and future-proof, allowing easy addition of new features, such as GenAI, while leveraging existing, widely available consumer technologies. Figure 1 presents a high-level overview of the system's components and their interactions.

\subsection{Overall Framework}
\tech's architecture is organized into key functional blocks, ensuring efficient information flow and decision-making. The core components include: \textbf{Vehicle Control}, \textbf{V2X Communication} and \textbf{GenAI Integration}. The system is designed to prioritize driving safety while enabling seamless integration with transportation infrastructure, laying the foundation for scalable, real-time CDA applications.

\subsection{Hardware Architecture Design}

\subsubsection{Edge Compute Platform}

We tested multiple AI edge devices, ultimately selecting the NVIDIA Jetson Orin Nano \cite{NVIDIAJetsonOrin} as the edge compute platform, offering a cost-effective yet powerful solution for running machine learning models and processing sensor data. This platform provides sufficient computational power to handle Level 2 ADAS tasks, including lane detection, obstacle avoidance, and real-time connectivity. It also hosts the edge LLM, which processes data locally for enhanced privacy and faster response times.

\subsubsection{Low Cost  Perception and Sensing}

A standard webcam \cite{logitech_c920} is integrated into the system to serve as the primary sensor for vision-based tasks such as object detection and environmental awareness. This affordable solution enables real-time image processing, essential for tasks like lane detection and obstacle recognition.

\subsubsection{Highly Interoperable Vehicle Interface}
The system interfaces with the vehicle’s onboard CAN bus through a CAN  adapter and the OBD-II port. This allows the system to access key vehicle data such as speed, steering angle, and braking force. The interface also enables sending control commands for steering, acceleration, and braking, creating a seamless integration with the vehicle’s existing driving systems. Integration with OpenDBC \cite{opendbc-git} enables broad compatibility across a variety of car makes and models, allowing for plug-and-play installation in supported vehicles.


    
\begin{figure}[H]
    \centering
    \includegraphics[width=0.9\linewidth]{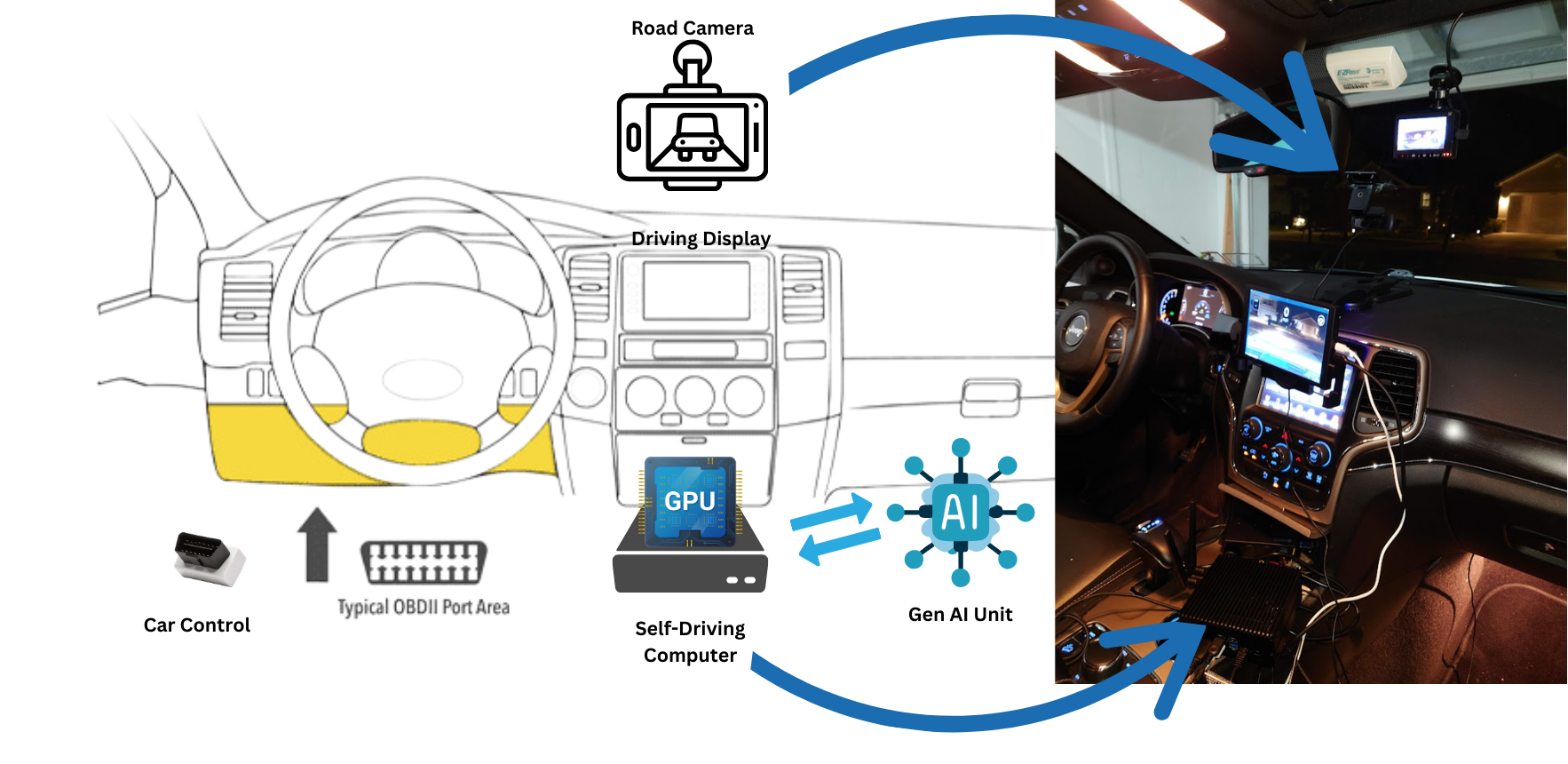}
    \caption{Communication links between hardware components}
    \label{fig:car_comm_layout}
\end{figure}

This combination of hardware components forms a cohesive and cost-effective experimental setup, with total Bill-of-Material(BOM), shown in table \ref{tab:hardware_cost}, costing less than \$1000 USD. This laying a solid foundation for developing and testing advanced autonomous driving features within an open-source framework. The communication links between hardware components are shown in figure \ref{fig:car_comm_layout}.

\begin{table*}[h!]
\centering
\begin{tabular}{|c|c|c|}
\hline
\textbf{Component} & \textbf{Description} & \textbf{Cost (USD)} \\
\hline
NVIDIA Jetson Orin Nano & Edge compute platform for processing ML models & \$480 \\
\hline
Standard Webcam (Logitech C920) & Vision sensor  & \$60\\
\hline
CAN Adapter & Interface for vehicle’s CAN bus to access vehicle data & \$100 \\
\hline
LTE and Wi-Fi Modules & Connectivity modules for V2X communication & \$250 \\
\hline
\end{tabular}
\caption{Cost of Hardware Components in AI-CDA4ALL System}
\label{tab:hardware_cost}
\end{table*}

\subsection{Software Architecture Design}

\subsubsection{Core Driving System}

The \tech software platform builds upon OpenPilot \cite{opendbc-git}, a widely adopted Level-2 open-source autonomous driving framework, to ensure affordability and compatibility across various vehicle models. OpenPilot provides a robust foundation for basic automated driving functions like adaptive cruise control and lane centering. By integrating open-source software, we ensure that the system is flexible and adaptable to new developments in autonomous driving technologies.

\subsubsection{Lightweight Connectivity Module}

Key to \tech's functionality is the integration of connectivity into the autonomy platform. Leveraging LTE and Wi-Fi communication protocols, the system establishes a Vehicle-to-Everything (V2X) network that supports real-time data exchange between vehicles, infrastructure, and cloud-based systems. This connectivity enhances safety and traffic efficiency by enabling the sharing of critical information, such as road conditions, traffic flow, and other dynamic factors that influence driving decisions.

Additionally, the use of open communication protocols such as MQTT ensures that the system can be easily integrated with other open-source platforms and future technologies, enabling ongoing development and collaboration with third-party developers, researchers, and vehicle manufacturers.

\subsubsection{GenAI Integration Module}

As generative AI technologies continue to advance, \tech incorporates local Large Language Models (LLMs) to enhance driving capabilities through reasoning, decision support, and real-time advisory systems. By deploying these models directly on the vehicle, the system ensures enhanced privacy, security, and responsiveness compared to traditional cloud-based solutions. This approach addresses the limitations of cloud connectivity, such as latency and dependence on stable internet access, ensuring that the system operates effectively even in remote areas or during network outages.

The integration of GenAI allows \tech to process real-time vehicle data, translate abstract reasoning into executable commands, and assist in vehicle control algorithms. For instance, LLMs can generate structured code that adapts to dynamic road conditions and offers personalized, context-aware driving suggestions. This integration represents a key innovation, bridging the gap between high-level AI capabilities and low-level vehicle control, facilitating smarter, safer driving experiences.

The GenAI software in \tech allows for seamless interaction between the vehicle and the driver using both voice and visual feedback. Through the camer's built-in microphone, the driver can activate the AI assistant using a wake word\cite{openwakeword-git} to initiate commands or ask questions. The AI can then respond through the vehicle’s speaker, providing verbal guidance or warnings related to driving conditions, such as traffic updates or safety alerts. Additionally, relevant information is displayed on the vehicle's screen, ensuring that drivers have both auditory and visual cues to enhance situational awareness and improve the driving experience.

\subsection{Middleware and Data Management}

Internally, the system uses middleware such as ZeroMQ for efficient communication between different software modules. This ensures that data from sensors, vehicle interfaces, V2X communications, and GenAI modules are seamlessly integrated and synchronized. Data logging and management tools ensure that the system can track its performance and provide insights for future improvements.

\subsection{Data Flow and Integration}

The flow of data through AI-CDA4ALL follows a structured process:

\begin{enumerate}
    \item 
    \emph{Standard Driving}: Camera input is processed by the perception module, which feeds into the planning module. The control module then adjusts vehicle actuation (steering, acceleration, braking).

    \item \emph{V2X Influence}: Infrastructure messages are received, parsed by the connectivity module, and influence planning or decision-making. For example, advisory speed limits could adjust the vehicle's speed.

    \item \emph{GenAI Interaction}: Sensor data and context are fed to the GenAI module, which generates recommendations. These recommendations are passed through the MetaAction API to modify vehicle settings, such as increasing the following distance.
\end{enumerate}


By integrating hardware, core driving software, connectivity, and GenAI modules, \tech provides a cohesive and efficient system for autonomous driving, emphasizing real-time processing and system reliability.

\section{Prototype Development and Demonstration}  

\subsection{Lightweight Connectivity}\label{connectivty}

\begin{figure}[h]
    \centering
    \includegraphics[width=0.6\linewidth]{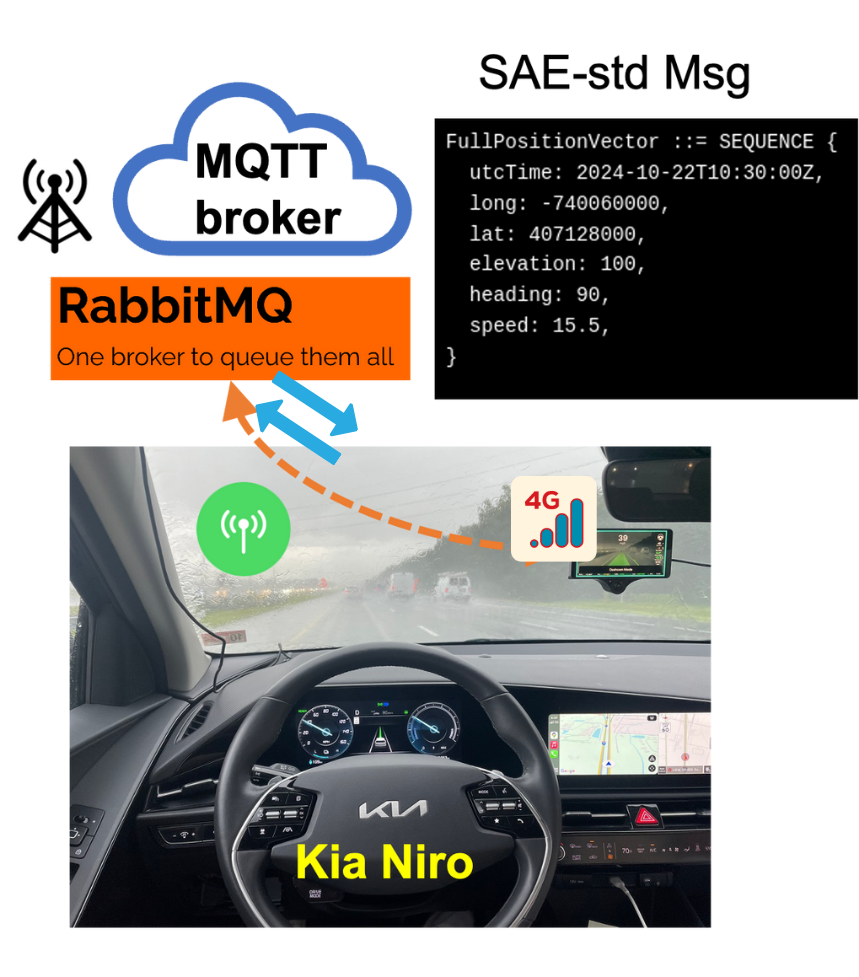}
    \caption{Data flow of MQTT communication using RabbitMQ}
    \label{fig:mqtt_rabbitmq}
\end{figure}

\tech has implemented a lightweight and efficient communication framework using the Message Queuing Telemetry Transport (MQTT) protocol. MQTT is a widespread communication protocol\cite{soni2017survey} for Internet of Things devices that facilitates seamless message exchange between connected edge devices, ensuring minimal latency and reliable communication even in constrained environments\cite{app9050848}. Each device in our system is connected to the a cloud MQTT server running RabbitMQ \cite{RabbitMQ} via either cellular LTE or WiFI, allowing for real-time data sharing across the network. Figure \ref{fig:mqtt_rabbitmq} depicted the workflow and architecture of \tech. This particularly advantageous for \tech, where timely and accurate information exchange is critical for enhancing road safety and traffic efficiency.

\begin{figure}[H]
    \centering
    \includegraphics[width=0.6
    \linewidth]{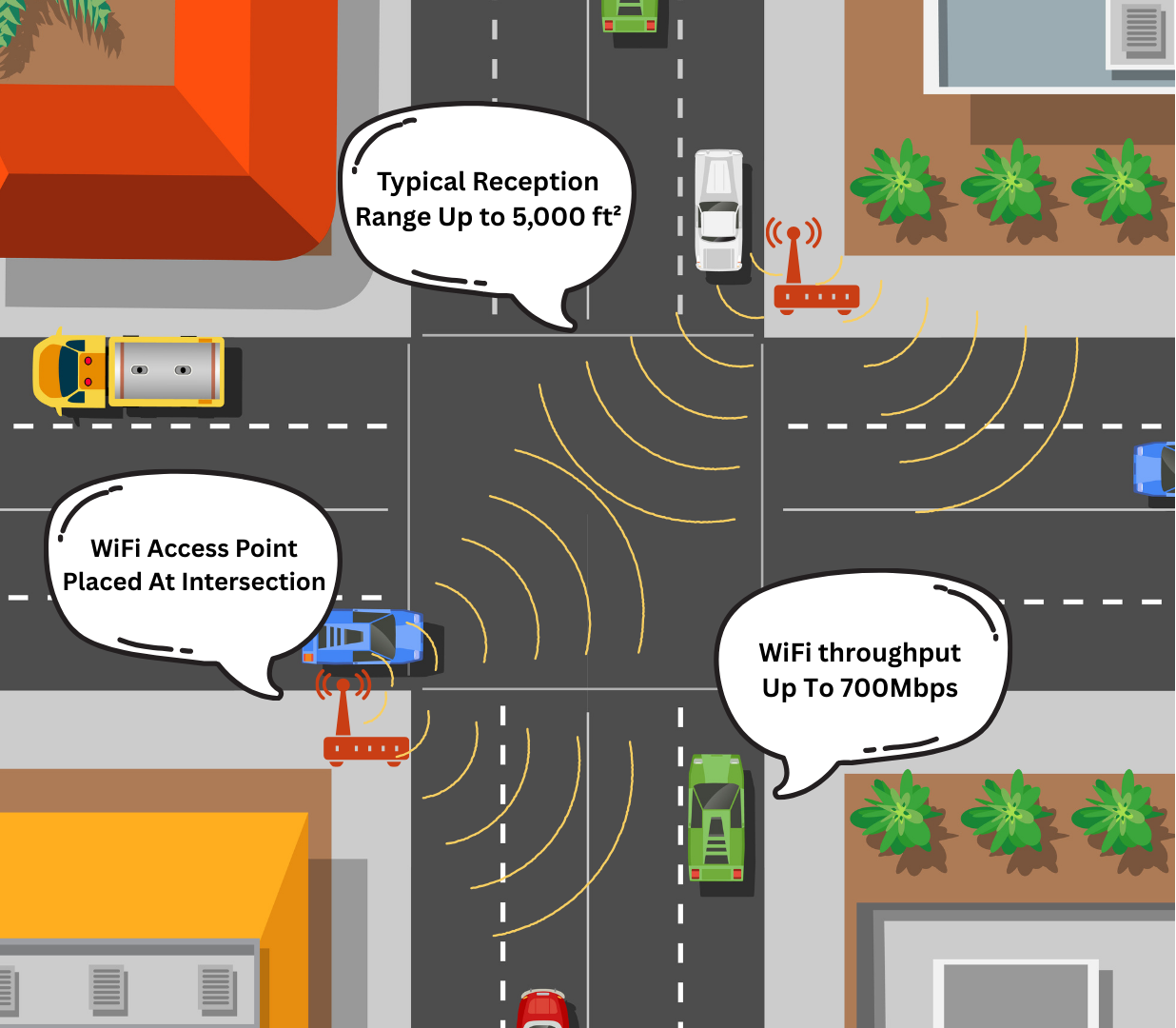}
    \caption{Example of Intersection WiFi Install Map}
    \label{fig:ai_comm}
\end{figure}

As the concept of the smart city continues to gain traction, the integration of WiFi plays a pivotal role in transforming transportation systems. WiFi, with its growing ubiquity in urban environments\cite{9965626}, can significantly reduce information latency by providing high-speed, low-cost communication between vehicles and infrastructure\cite{smartcities6020048}. Figure \ref{fig:ai_comm} shown a reference design of a smart city concept with WiFi access point at intersection act as Road-Side Unit (RSU). This not only accelerates data transmission but also supports high-bandwidth applications like real-time sensor data streaming and vehicle-to-vehicle (V2V) communication. The reduced latency improves response times for autonomous driving systems, enabling quicker decision-making and enhancing overall driving safety.

In addition, the inclusion of LTE networks ensures greater coverage, particularly in areas where WiFi connectivity may be sparse. LTE’s relatively wide reach combined with its low latency makes it an ideal candidate for Vehicle-to-Everything (V2X) communications, extending beyond vehicles to include traffic lights, road sensors, and other infrastructure components. This enhances the collaborative awareness of the entire transportation system, allowing for dynamic adjustments based on real-time traffic data and improving overall traffic flow.

The use of MQTT, combined with WiFi and LTE communication protocols, is a highly practical solution for modern transportation needs. It reduces infrastructure complexity, provides scalability, and supports the efficient exchange of critical data, all while maintaining low operational costs. By focusing on lightweight connectivity, our approach makes it possible to deploy advanced autonomous driving systems at scale, making \tech more accessible and affordable, while supporting the evolution of smart, connected cities.


\subsection{Transmission of Connectivity Message}

\tech includes the development of an SAE standard V2X message - J2735\cite{j2735}. Our library that can encode SAE standard messages, ensuring consistency and compatibility with established transportation protocols. These messages can then be transmitted over Wi-Fi using the QUIC protocol \cite{rfc9000}, a high-performance transport layer protocol known for its low-latency and secure data transmission. This solution enhances real-time communication and data exchange between vehicles and infrastructure, enabling efficient, scalable, and secure transmission of SAE-standard messages in connected and automated transportation systems.

By using WiFi Network connection, (demonstrated in \ref{compare_wifi}) that WiFi 6 (5GHz), offers low latency ($<$ 10ms), high-speed (up to 4.3 Gbps), and substantial coverage (up to 300 meters), making it an ideal connectivity solution for non-safety-critical applications. The speed and latency of WiFi 6 are significantly better than those of traditional 2.4GHz WiFi routers and LTE connections.

In comparison, LTE connectivity, although reliable, has higher latency ($<$ 100ms) and offers variable speeds based on network congestion thus making them less suitable for applications that require higher data rates and wider coverage.

Thus, new WiFi standard such as WiFi 6 adapoted by \tech offers a highly effective, scalable, and fast solution for applications that don't require immediate response times, positioning it as a practical connectivity option for non-safety-critical use cases.

\begin{table*}[htbp]
\begin{tabular}{|c|c|c|c|}
\hline
\textbf{Metric} & \textbf{WiFi 6 (5GHz) - \tech RSU \cite{u7outdoorspec}} & \textbf{WiFi 4/5 (2.4GHz) - Regular Router} & \textbf{Cellular LTE} \\
\hline 
Latency & $< 10$ ms & $< 50$ ms & $< 100$ ms \\
\hline 
Speed & Up to 4.3 Gbps & Up to 100 Mbps & $\approx$50Mbps (varies with network condition) \\
\hline 
Coverage & Up to 465 meters & Up to 100 meters & Dependent on cell tower range \\
\hline
\end{tabular}
\caption{WiFi and LTE transmission comparison}
\label{compare_wifi}
\end{table*}

\subsection{Integration of GenAI} \label{local_gen_ai}

As generative AI and Large Language Model (LLMs) become widely adapted in varies industry, \tech has leveraged the use of local generative AI models to assist drivers to both visually and audibly show advisory information and assist the control algorithm, this presents significant opportunities for improving road safety, traffic management, and vehicle autonomy. By deploying local LLMs directly on vehicles, this approach offers enhanced privacy, security, and real-time responsiveness—addressing the critical challenges faced by traditional cloud-based solutions.

A local LLM processes data entirely on the vehicle, ensuring that sensitive information such as CAN data, driver behavior, and real-time telemetry remains secure. Compared to Cloud-based models which introduce network latency, which is unacceptable in safety-critical applications such as lane departure avoidance. It also heavily relies on stable internet connectivity, which may not be available in remote areas or during network outages. In contrast, local LLMs can function independently, ensuring continuous operation even in offline environments. As tested with an NVIDIA Jetson AGX Orin, we achieved a total request-to-response time of mean of 5.07 seconds and median of 4.61 seconds with LLaMA 3.2 model.

\begin{figure*}[H]
    \centering
    \includegraphics[width=0.3\linewidth]{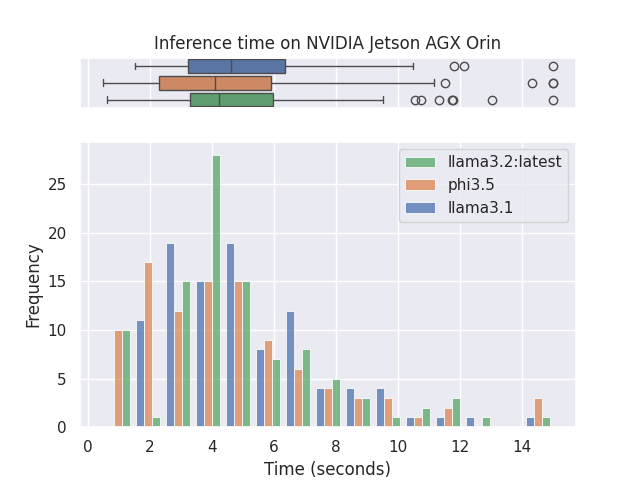}
    \includegraphics[width=0.3\linewidth]{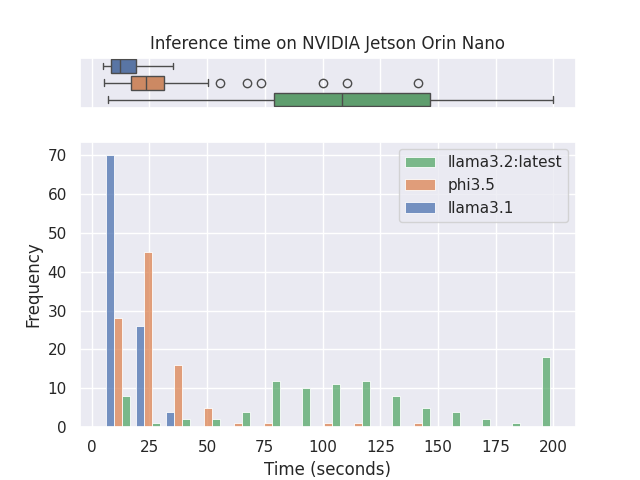}
    \includegraphics[width=0.3\linewidth]{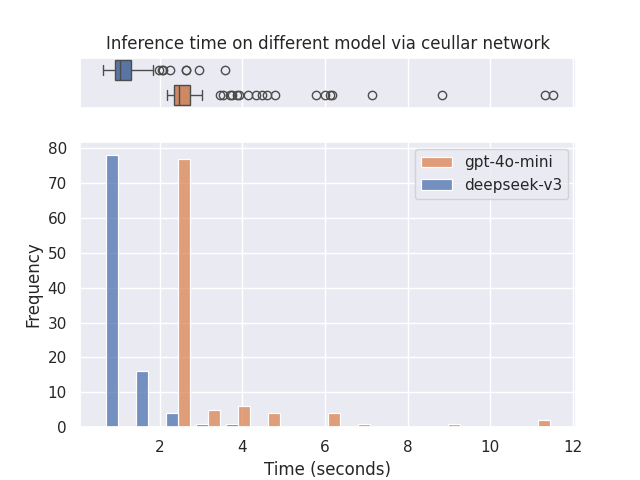}
    \caption{Inference time on NVIDIA AGX Orin \& Orin Nano \& Cloud LLM}
    \label{fig:infernce_time_orin}
\end{figure*}

\begin{table}[htbp]
\begin{tabular}{|c|c|c|c|c|}
\hline
Model & Runner & $\overline{T_{net}}$ & $Med \ T_{net}$  & $\overline{L_{token}}$ \\
\hline
LLaMA 3.2 & AGX Orin & 5.07 & 4.61 & 185 \\
\hline
LLaMA 3.1 & AGX Orin & 4.86 & 4.22 & 106 \\
\hline
Phi 3.5 & AGX Orin & 4.62 & 4.09 & 279 \\
\hline
LLaMA 3.2 & Orin Nano & 14.10 & 12.20 & 202 \\
\hline
LLaMA 3.1 & Orin Nano & 112.44 & 108.27 & 119 \\
\hline
Phi3.5 & Orin Nano & 27.53 & 23.54 & 468 \\
\hline
GPT-4o-mini & Cloud via LTE & 1.20 & 1.05 & 42 \\
\hline
DeepSeek3-R1 & Cloud via LTE & 3.07 & 2.48 & 54 \\
\hline
\end{tabular}
\caption{Model speed and average token}
\label{model_compare}
\end{table}

\begin{figure}[H]
    \centering
    \includegraphics[width=0.9\linewidth]{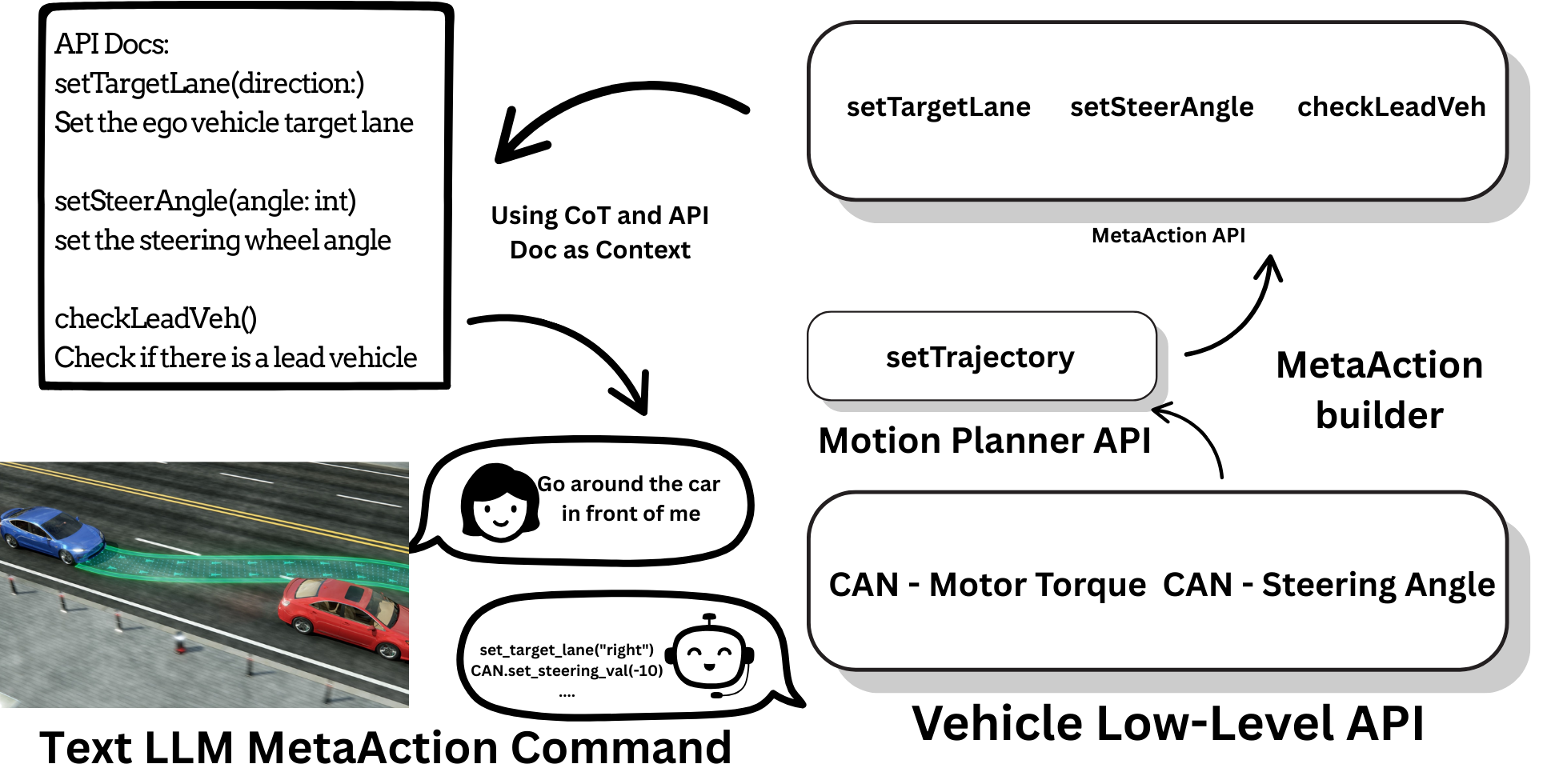}
    \caption{MetaAction Based LLM Command Workflow}
    \label{fig:metaaction_workflow}
\end{figure}

The local LLM can also act as a code generation agent, translating Chain-of-Thought (CoT) reasoning into real control commands for the vehicle via the rest of \tech's control API. This capability allows the LLM to dynamically adapt its responses based on real-time data and execute common vehicle maneuvers. We proposed a MetaAction based LLM command shown in figure \ref{fig:metaaction_workflow}, that takes advantage of low level control API while exposing only high level commands that LLM can easily work with. This interface not only can benefit LLM generation but also benefit any third party integration or low-code development on vehicles.

Since \tech has access to low level vehicle data and controls, a driving assistant can be created to send control command to the low level motion controller using API specified by the prompt. 

\begin{figure}[h]
    \centering
    \includegraphics[width=0.9\linewidth]{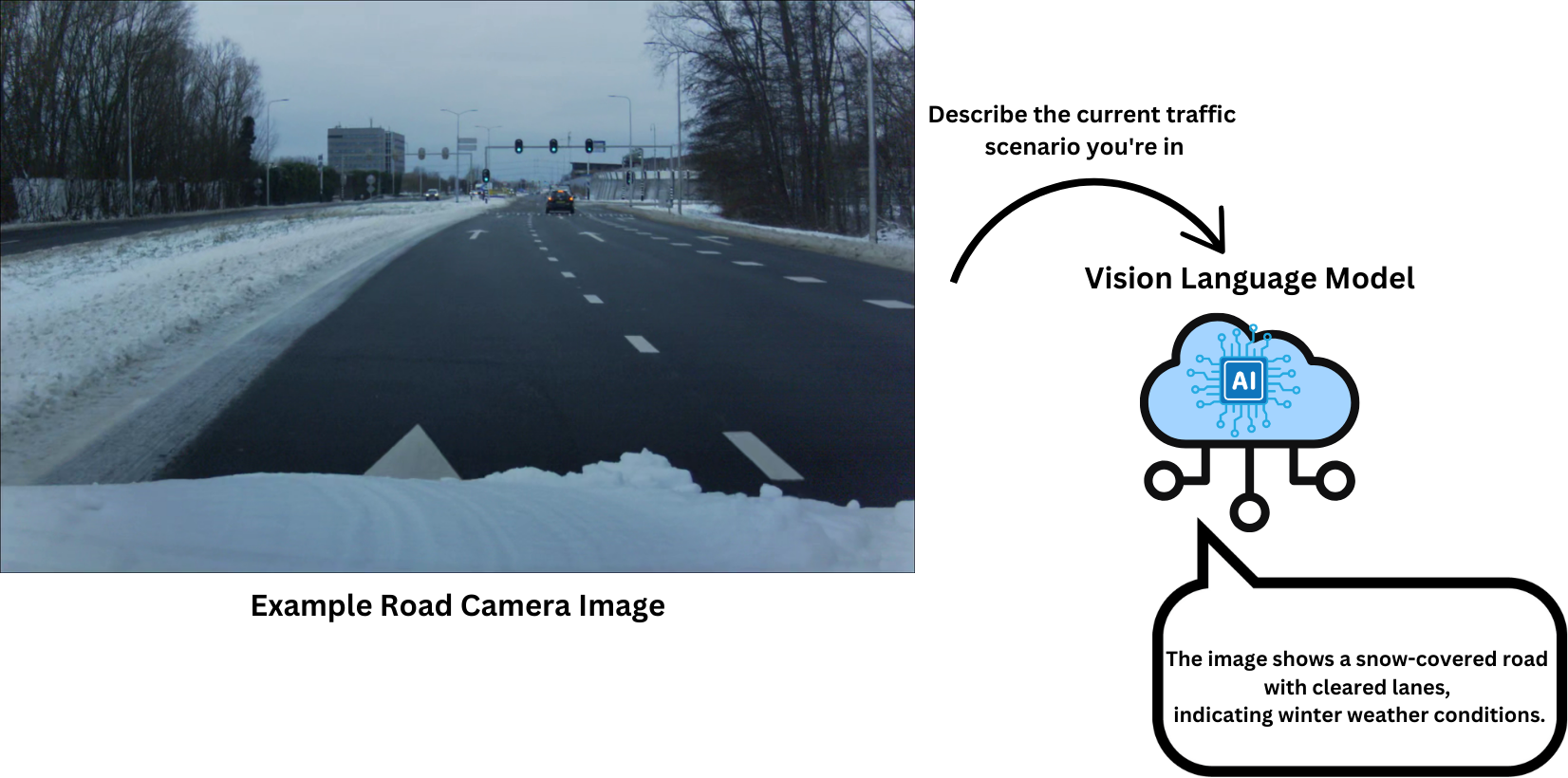}
    \caption{Example of Vision Language Model (GPT4o-mini) advisory}
    \label{fig:vlm_adviosry}
\end{figure}

As vision language model gains in popularity, more advisory information can be given on road condition and surrounding environments. Its ability to recognize the time of day,  current weather conditions, and its proficiency in recognizing and interpreting
traffic lights and sign all hold paramount significance in shaping the autonomous
driving system’s decision-making process \cite{wen2023roadgpt4visionearlyexplorations} shown in figure \ref{fig:vlm_adviosry}. 

\section{Application and Real-world Use Case}

\subsection{High-quality driving data collection}

Since \tech is based on the popular open-source ADAS system OpenPilot, \tech has modified to suits for a general purpose, high quality data collection platform. While OpenPilot is intended to use with a compatible vehicle, \tech allows any vehicle to use its powerful lane detection model and its's accompanying hardware Comma 3X's CAN adapter function to collect any vehicle data from the vehicle using the OBD port. \tech also have a suites of post-processing tools designed for extracting useful information such as engine fuel efficiency for traditional ICE vehicle or energy consumption for EV.

Our group's previous work - OpenLKA \cite{wang2025openlkaopendatasetlane} has already leveraged this framework to collect a high quality dataset with a variety of vehicle lane keep assist functionality under different conditions as a testing ground to assist with vehicle autonomy research.

\subsection{Online traffic data acquisition and advisory speeds}

As many transportation agencies continue to integrate connectivity solutions into their infrastructures, not many road users are aware the benefit of using such information. \tech can be used to seemless integrate information from DOT and other transportation agencies. 

\begin{figure}[h]
    \centering
    \includegraphics[width=0.9\linewidth]{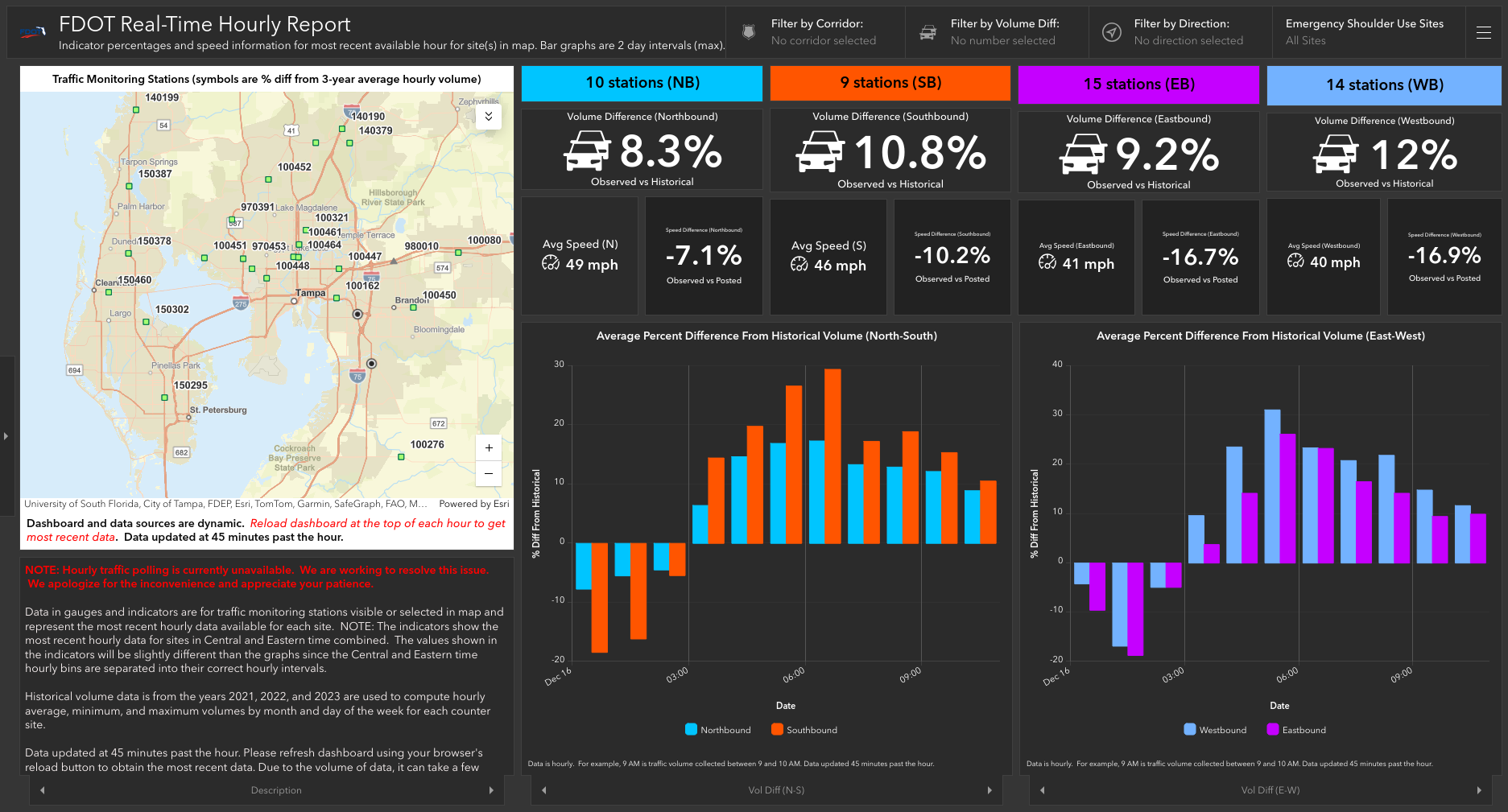}
    \caption{Screenshot of FDOT Traffic Report}
    \label{fig:fdot_traffic}
\end{figure}

As a demonstration, we used the Florida DOT 511 website to crawl data and be displayed on the \tech interface shown in figure \ref{fig:fdot_traffic} We used the data and displayed a traffic dashboard which shows the current traffic patterns in the Tampa Bay area.

As described before in section \ref{connectivty}, DOT and other transportation agencies may push a advisory speed and using methods we proposed to command the control flow to the vehicles along a specific route to reduce congestion or enhance safety.

To command a platoon or a single \tech equipped vehicle along a specific route, the command center would send out advise speed value to the central server. \tech would receive such values in real-time where the devices are connected to the internet and based on their location, apply the advisory speed value to the control loop of the lateral control algorithm. 

\begin{figure}[h]
    \centering
    \includegraphics[width=0.9\linewidth]{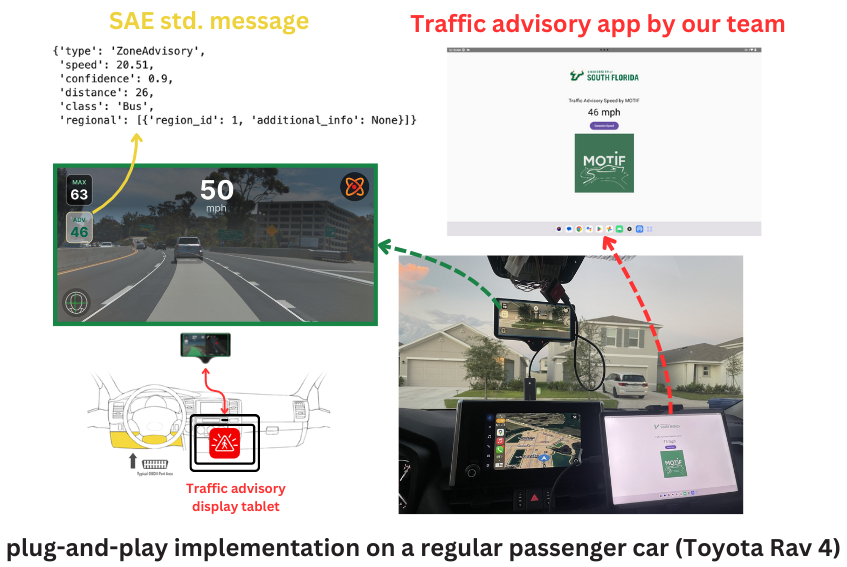}
    \caption{Advisory Speed Application In a Toyota RAV4}
    \label{fig:advisory_demo}
\end{figure}

We have successfully packaged the advisory speed app to be compatible with most mobile phone and tablet running Apple iOS and Android in figure \ref{fig:advisory_demo}, in future development, we hope to bring these displays to non-\tech equpied vehicle such that the information such as advisory speed would show thru the vehicle infotainment system such as CarPlay and Android Auto. For transportation agencies wanting a fast adaption rate, \tech lowered the barrier to retro-fit almost all vehicle with access with a compatible vehicle infotainment system.

\subsection{Transportation Agencies Adaption}

\begin{figure}[h]
    \centering
    \includegraphics[width=0.9\linewidth]{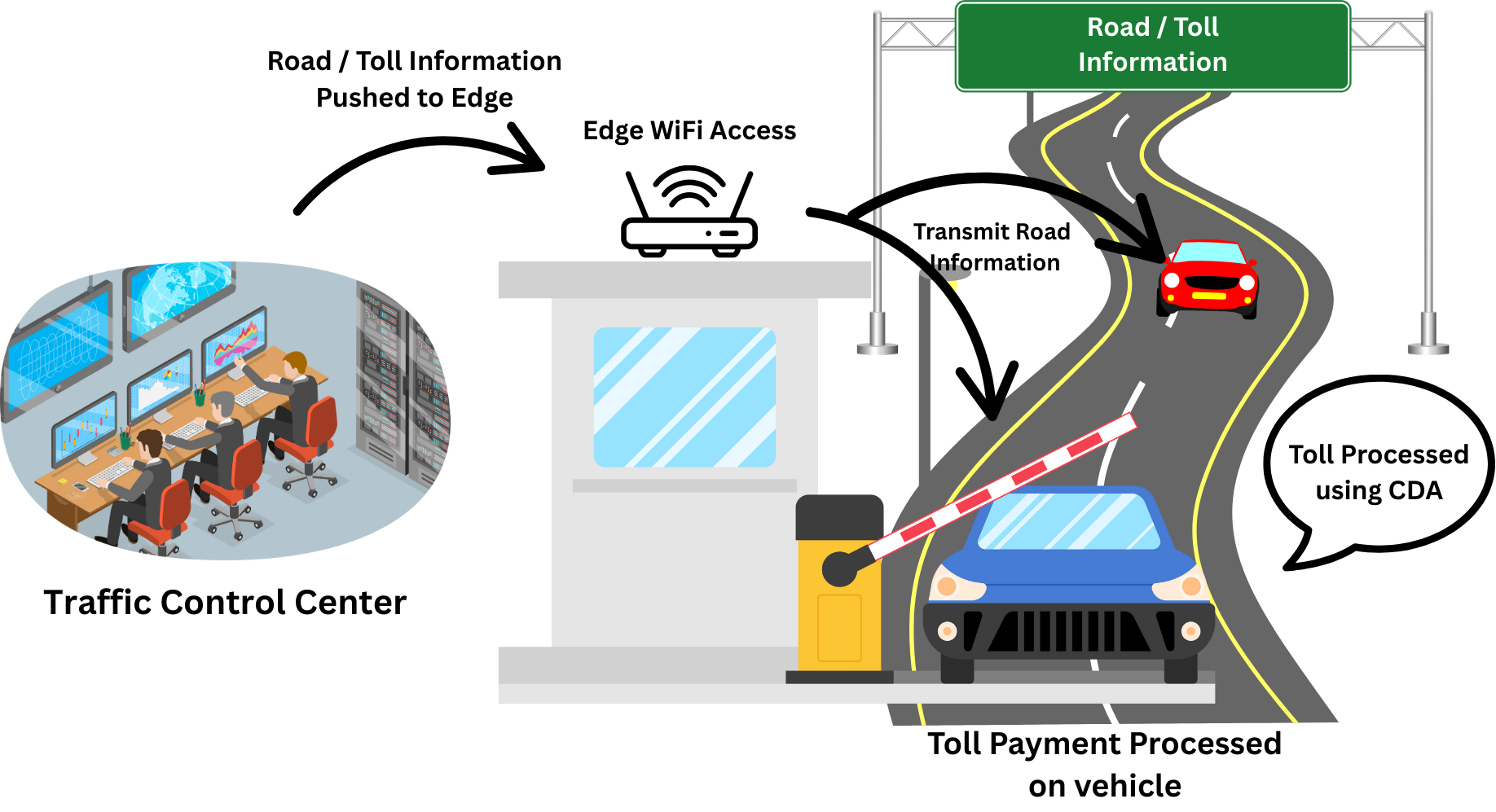}
    \caption{Example of Toll Road Use case}
    \label{fig:connected_toll}
\end{figure}

\tech addresses a critical gap in CAV research by providing a scalable, cost-effective retrofit solution for legacy vehicles. While most current research focuses on deploying CAV technologies in new vehicles\, our framework demonstrates that retrofitting existing fleets can achieve similar results at a fraction of the cost. By leveraging lightweight connectivity protocols, edge computing, and open-source platforms, \tech enables real-world testing of V2X communication and autonomous functionalities in diverse environments. Agencies such as toll authority can leverage \tech to minimize cost on toll collection infrastructure when comparing to traditional RFID toll tag system. The system can also send near real-time traffic alert to road which can greatly improve road user experience shown in figure \ref{fig:connected_toll}.

\bibliographystyle{IEEEtran}  
\bibliography{references}

\end{document}